# Thick $Bi_2Sr_2CaCu_2O_{8+\delta}$ films grown by liquid-phase epitaxy for Josephson THz applications

Y Simsek[1], V Vlasko-Vlasov[1], A E Koshelev[1], T Benseman[1,2], Y Hao[1,3] I Kesgin[4], H Claus[1], J Pearson[1], W-K Kwok[1] and U Welp[1]

[1] *Materials Science Division, Argonne National Laboratory, Argonne, IL 60439, USA*

[2] *Department of Physics, CUNY Queens College, Queens, NY 11367, USA*

[3] *Department of Physics, University of Illinois at Chicago, Chicago, IL 60607, USA*

[4] *Advanced Photon Source, Argonne National Laboratory, Argonne, IL 60439, USA*

Email: simsek@anl.gov

## Abstract

Theoretical and experimental studies of intrinsic Josephson junctions that naturally occur in high-$T_c$ superconducting $Bi_2Sr_2CaCu_2O_{8+\delta}$ (Bi-2212) have demonstrated their potential for novel types of compact devices for the generation and sensing of electromagnetic radiation in the THz range. Here, we show that the THz-on-a-chip concept may be realized in liquid phase epitaxial-grown (LPE) thick Bi-2212 films. We have grown μm-thick Bi-2212 LPE films on MgO substrates. These films display excellent *c*-axis alignment and single crystal grains of about 650 × 150 μm² in size. A branched current-voltage characteristic was clearly observed in *c*-axis transport, which is a clear signature of underdamped intrinsic Josephson junctions, and a prerequisite for THz-generation. We discuss LPE growth conditions allowing improvement of the structural quality and superconducting properties of Bi-2212 films for THz applications.

## 1. Introduction

Intrinsic Josephson junctions (IJJs) [1] occur naturally in the highly anisotropic layered high-temperature superconductor (HTS) $Bi_2Sr_2CaCu_2O_{8+\delta}$ (Bi-2212). They exhibit the fundamental phenomena associated with the Josephson effect, namely resonant plasma oscillations [2], Shapiro steps [3] and Fiske resonances [4] in the frequency range of 0.3 - 1.3 THz which has been difficult to cover with conventional solid-state devices [5, 6]. More recently, oscillatory behavior up to 13 THz has been reported for small BSCCO structures [7, 8]. These properties allow IJJs to act as compact solid-state generators and detectors for electromagnetic waves in this so-called THz-gap for various emerging applications in materials spectroscopy, medical diagnostics, high-bandwidth communications and security [6]. An ideal emitter for practical applications should produce a powerful and tunable THz output. The linear relationship between emission frequency and applied voltage per junction as expressed by the AC Josephson relation – and the cavity mode's dependence on the in-plane superconducting penetration depth – enable tunable THz radiation spectra by changing the bias voltage and/or sample temperature [9, 10, 11]. Furthermore, emission power scales as the square of the number of IJJs synchronized in a resonant mode [2, 12] and can thus be enhanced to technologically useful levels by changing the geometry of the resonator arrays. In fact, coherent THz radiation with power as high as 0.6 mW was experimentally demonstrated by synchronizing three resonators through propagation of Josephson plasma waves in the Bi-2212 base crystal [13]. This versatile platform can also transmit THz waves through Bi-2212 crystals with low losses [14]. A major step in realizing THz-on-a-chip systems, that is, a single chip that combines and integrates devices such as emitters working in the THz range, high-frequency detectors, waveguides, high-bandwidth receivers, all operating on the basis of Josephson plasma wave (JPW), is the synthesis of large-area epitaxial thick films of Bi-2212. The 2D platform and highly thermally conductive MgO substrate can also provide a uniform heat sink for the compact THz devices. Here, we demonstrate that liquid phase epitaxial (LPE) grown thick Bi-2212 films may be a candidate platform to realize THz-on-a-chip devices.

The $Bi_2Sr_2Ca_{n-1}Cu_nO_{2n+4+\delta}$ high-$T_c$ parent material (BSCCO) has three well-known superconducting phases; ($n$ = 1) $Bi_2Sr_2CuO_{6+\delta}$ (Bi-2201) with $T_c \leq$ 20 K, ($n$ = 2) $Bi_2Sr_2CaCu_2O_{8+\delta}$ (Bi-2212) with $T_c \leq$ 90 K and ($n$ = 3) $Bi_2Sr_2Ca_2Cu_3O_{10+\delta}$ (Bi-2223) with $T_c \leq$ 110K [15, 16]. Here, $n$ is the number of superconducting Cu-O layers that are separated by Ca layer(s) in Bi-2212 and Bi-2223. In these phases, the superconducting layers are separated by insulating Bi-O and Sr-O layers resulting in natural stacks of intrinsic Josephson junctions. Since defects, impurities and misalignment in these complex crystallographic layered structures degrade the superconducting properties of the material, uniform and single-phase crystallographic film structure with well-aligned multi-cation oxide layers on a substrate is required for high-performance applications. For the practical implementation and scale-up of superconducting THz technology, high-quality and thick epitaxial Bi-2212 films with a large-size grain texture are required.

Vapour phase methods such as sputtering, pulse laser deposition (PLD) and molecular beam epitaxy (MBE) allow for the growth of epitaxial high-$T_c$ films. However, these methods are slow and not well suited for the growth of µm-thick films. [17, 18]. Alternatively, solid/liquid phase methods enable the growth of thick films. In solid-phase epitaxy, the crystallization of pre-deposited amorphous high-$T_c$ components can be driven on a suitable substrate in the presence of a temperature gradient [19]. However, the complete transformation of a thick amorphous precursor film into a crystalline single phase is challenging in the solid phase. In a variation, a layer of precursor material is confined between two substrates and molten and re-solidified in an infra-red image surface [20, 21, 22]. After splitting the substrates apart, epitaxial superconducting Bi-2212 films were obtained, albeit with irregular and uncontrolled thickness. Encouragingly, however, this two-dimensional variation of the Bridgman technique did result in the successful growth of stacked IJJs, from which THz emission has been demonstrated [22].

Liquid-phase epitaxial (LPE) growth of high-$T_c$ superconductor films is performed in a saturated solution at temperatures below the melting point of the BSCCO precursor composition. In the liquid medium, the dissolved components have high mobility and can access the correct nucleation positions, enabling the self-assembly of high-quality crystalline films on the substrate [23, 24]. The LPE technique is a promising approach for growing wafer-scale, thick and single-crystalline high-$T_c$ superconducting films [25, 26, 27]. In this paper we show experimentally that given the correct growth parameters, LPE can produce Bi-2212 IJJs with characteristics comparable with those of a single crystal and large grain size, while avoiding the need to cleave two substrates apart.

Building on the procedures outlined in [25], we implemented an optimized LPE system through variation of the growth conditions. We demonstrate the high quality of our LPE-grown films through structural and stoichiometric characterizations as well as transport and magnetization measurements of their superconducting properties. The films display excellent $c$-axis alignment and single crystal grains of about $650 \times 150$ μm$^2$ in size, which is large enough for the fabrication of THz-emitting devices. A branched current-voltage characteristic was clearly observed, which is a clear signature of underdamped intrinsic Josephson junctions, and a prerequisite for THz-generation.

## 2. Experimental

Figure 1 shows an illustration of our LPE setup incorporated into a vertical furnace. The LPE growth mechanism is carried out in a 110 mL platinum (Pt) crucible containing BSCCO precursor components and a large amount of KCl flux maintained at a high temperature. The crucible is placed on top of a mobile alumina stage located at the center of the heat zone of the furnace. This position provides a desired downward temperature gradient to drive the dissolved BSCCO components in the molten KCl flux from the bottom to the top of the crucible. A substrate, clamped onto a Pt holder, is placed just below the liquid surface. The upward chemical transport of the BSCCO solute in KCl leads to its deposition onto the substrate and consequently to the formation of a crystalline film. The synthesis temperature is monitored by an $R$-type thermocouple inserted into the crucible at the same height as the substrate (see inset picture in Figure 1). A second thermocouple connected to the temperature control unit of the furnace is used to stabilize the temperature in the hot zone. In order to avoid temperature fluctuations caused by air flow the bottom and top openings of the furnace are completely sealed during the growth process.

The precursor composition (20 g) was prepared using the following procedure: (i) Bi$_2$O$_3$, SrCO$_3$, CuCO$_3$, and CuO components were mixed with a molar ratio corresponding to Bi-2223, grounded in a mortar and loaded into the Pt crucible. (As will be discussed later, excess amounts of calcium and copper (i.e. 2223-composition) are in fact crucial for the growth of good quality Bi-2212 films.) (ii) The precursors heated to 850°C with a ramp rate of 5°C/minute and calcinated for 1 hour at the center of the furnace's hot zone, then rapidly quenched in air to room temperature. This resulted in partially molten precursor material stuck to the bottom of the crucible. We then added 100 g of KCl flux to the crucible, heated it to 800°C and maintained this temperature for 1 hour, promoting the melting of the KCl flux. Subsequently, the crucible was quickly removed from the furnace and quenched in air to room temperature. (iii) In the final step, the crucible, covered with an alumina lid, was placed again at the center of the heat zone and soaked at the growth temperature for 10 hours to obtain a supersaturated BSCCO solution. After removing the lid from the crucible, the Pt holder with the clamped substrate was slowly lowered into the hot zone and immersed into the crucible. The substrate was kept approximately 8-10 mm below the liquid surface. At the end of the film growth, the substrate is lifted out of the solution and the residual KCl flux on the substrate is immediately removed by spinning the holder at 400 rpm. The holder is slowly extracted from the furnace, cooled in air, and any residual KCl on the film is dissolved in deionized water.

Here we describe results obtained on four epitaxial Bi-2212 films on MgO (100) substrates sequentially grown from the same Bi-2223 precursor using the LPE process each performed at 870°C for 10 hours. We discovered that excess amount of calcium and copper (2223-composition) to be crucial for the growth of good quality Bi-2212 films. In all growth runs, the temperature gradient of the solution, 1 cm below the substrate was around 2-3°C/cm. Before immersing the substrate into the solution, we performed the following steps: (i) crust layers on the top of the liquid arising due to the evaporation of KCl was removed with a Pt hook, (ii) 5 g of KCl was added to compensate for the evaporated KCl and (iii) the replenished solution was soaked at 870°C for 10 hours using the alumina lid on the top of the crucible to re-establish the supersaturated solution as described above.

The crystalline structure and atomic composition of the Bi-2212 films were studied by X-ray diffraction (XRD) and energy-dispersive X-ray spectroscopy (EDS) respectively while their surface morphology was characterized by scanning electron microscopy (SEM) and polarized-light optical microscopy. The superconducting properties of the resulting films were tested by magnetization and by in-plane and out-of-plane transport measurements.

### 3. Results and discussion

The deposition rate of the LPE-grown Bi-2212 film is comparatively high, enabling the growth of a three-micron thick film in ~20 hours. Obviously, this rate changes with growth parameters such as temperature, temperature gradient, saturation of the solution and heat convection in the crucible. Indeed, maintaining stable conditions of these parameters during the growth process is essential to achieve consistent single-phase crystallinity across the film thickness. However, effects due to the evaporation of KCl can perturb the stable growth condition in the crucible. Since vapourization induces a cooler liquid surface, the BSCCO precursor transport is diverted from the substrate to the liquid surface [25]. This causes the formation of a crust on the liquid surface, composed of Bi-2212 crystal flakes and amorphous BSCCO precursor, which may alter the growth dynamics. The enhanced convective transport inside the crucible also affects the film growth on the substrate since it promotes the nucleation of misaligned grains and the deposition of amorphous material on top of the Bi-2212 film. We were able to significantly reduce the KCl evaporation by placing the crucible exactly at the center of the hot zone thereby reducing any temperature gradients, and by placing a lid on the crucible. In addition, to avoid attachment of any crust on the substrate during the initial growth process, we spun the substrate holder at 90 rpm while immersing it into the solution.

In LPE film growth, the substrates must be selected carefully. In addition to the usual requirements such as crystallographic lattice match, thermodynamic compatibility, surface quality, etc., the substrate must be chemically resistant to the solution. In our study, we carried out LPE growth of Bi-2212 films on a variety of $1 \times 1$ cm$^2$ substrates (one-side polished surface) such as $NdGaO_3$ (001) $SrTiO_3$ (100) MgO (100) and YSZ (100) (yttria-stabilized zirconia). However, we could grow high-quality Bi-2212 films only on MgO substrates, as other substrates were not chemically resistant to the aggressive growth solution at 870°C. For instance, our EDS results revealed that Ga atoms in $NdGaO_3$ substrates diffuse into the film structure above 850°C and degrade the superconducting properties. Besides the substrates, all materials in contact with the solution must be chemically inert. In preliminary experiments performed in Pt-Rh%10 crucibles, we observed unexpected corrosion of the crucible walls, and a small amount of rhodium (Rh) was detected in the EDS characterization of the Bi-2212 films. This corrosion may originate from Rh based compounds in the crucible alloy which are prone to melt below the melting point of pure Rh. Recently, we carried out a large number of LPE growth experiments in a pure Pt crucible. No visible corrosion on the crucible walls nor any Pt impurities in the grown films were observed that could originate from alloying effects [28].

Figure 2 shows optical and SEM pictures of a Bi-2212 film grown on an MgO substrate. This is the third film grown from the same LPE growth batch. The film is ~1 μm thick and is free of voids, (see Figure 2a). The texture of the film is characterized by large-sized grains with some in-plane misorientation and with very well aligned *c*-axes. Polarized light microscopy (see Figure 2b) images show that single-crystalline grains reach sizes of 630 × 150 μm$^2$. This area is large enough to fabricate Bi-2212 resonators for THz-emission [2, 10, 13]. The SEM image in Figure 2c shows a smooth surface morphology on the scale of tens of micrometers.

The structural characterization of the grown Bi-2212 films was carried out by X-ray diffraction with Cu Kα radiation. θ - 2θ scans (see Figure 3) of the four sequentially grown films (BF-1, BF-2, BF-3 and BF-4) reveal sharp and intense (00ℓ) peaks, indicating uniform *c*-axis alignment of crystalline grains on the substrates. The XRD scans of the first three films (BF-1, BF-2, BF-3) depict mostly diffraction peaks corresponding to the Bi-2212 phase, while reflection peaks corresponding to both Bi-2212 and Bi-2201 phases were clearly observed in the XRD results of the last grown film (BF-4). As the peaks from the Bi-2201 phase in our films are weak, X-ray intensities from these films are plotted on a logarithmic scale in Figure 3a. The large intensity difference between the Bi-2212 and Bi-2201 reflections implies a volume fraction of Bi-2201 phase of ~3% in the film BF-4. Moreover tiny Bi-2201 reflection peaks are visible on the XRD data of these films BF-1 and BF-3 (note the logarithmic scale), but their Bi-2201 volume fractions are negligible relative to Bi-2212 phase. The high crystalline quality of the epitaxial films was also confirmed by X-ray rocking curve measurements about the most intense diffraction peaks. Figure 3b shows rocking curves of (00$\underline{10}$) and (00$\underline{12}$) XRD peaks recorded from two films (BF-2 and BF-3). The sharp single peaks with narrow full width at half maximum (FWHM) (0.06° and 0.07° of the two films, respectively) indicate the pure and perfect in-plane crystalline mosaic spread of the epitaxial films.

All EDS analyses were conducted on a large area (250 × 250 μm$^2$) to detect the stoichiometric atomic ratios and stoichiometric uniformity of each crystalline film. In Figure 4 we show the atomic percentages of Bi, Sr, Ca and Cu detected on two different regions on each film. Dashed lines indicate the theoretical elemental ratios calculated with respect to the chemical stoichiometry of single phase Bi-2212. As shown in Figure 4, the EDS results for the first three films exhibit atomic ratios close to the stoichiometry of Bi-2212. The Ca content in film BF-4 is substantially lower than the nominal stoichiometry of the Bi-2212 phase. This originates not only from the presence of the Bi-2201 phase in the film, but also from a large number of Ca vacancies in the Bi-2212 phase. Typically, the Ca concentration decreases with the synthesis time of the LPE growth material, which directly influences the stoichiometry of the resulting films. As discussed by Balestrino *et al.* [29], the Ca-deficiency may originate from the lower solubility relative to Bi and Sr in the growth solution. For example, Ca, as well as Cu, is believed to form large complex aggregates in the BSCCO melt with time, which accumulate mostly at the bottom of the crucible. For this reason, a Ca/Cu-rich starting composition, i.e., Bi-2223, is more likely to yield high-quality Bi-2212 LPE-films.

SQUID magnetometry was used to determine the critical temperature ($T_c$) and transition width ($\Delta T_c$) of the films. Since some superconducting parameters such as $T_c$, critical current density ($J_c$), and energy gap ($2\Delta$) are very sensitive to small variations in the Bi-2212 stoichiometry, magnetometry results also provide a general indication of crystallographic film uniformity. All as-grown films were annealed in an oxygen atmosphere at 550°C for 15 hours to investigate variations of their superconducting transitions with oxygen stoichiometry. Figure 5a shows the temperature dependence of the magnetic moment of the annealed films measured on warming in a field of 0.1 G after zero-field cooling. The transition curves of films BF-1, BF-2 and BF-3 are relatively sharp, while the last grown film, (BF-4) exhibits a broad transition extending from 80 K to below 40 K. These observations are consistent with the XRD and EDS results presented above. Moreover, $T_c \sim 76$ K of film BF-1 seen in Figure 5a is slightly lower than that of BF-2 and BF-3, $T_c \sim 80$ K, which is most likely a consequence of the slightly reduced Cu ratio found in this film (see Figure 4). Figure 5b shows a comparison of the temperature dependence of the in-plane resistivity measured on as-grown and annealed sections of film BF-3. The resistance was measured with the conventional 4-probe technique on pieces of approximately $1 \times 3$ mm$^2$ size. As seen in Figure 5b, upon annealing in oxygen, the width of the superconducting transition and the normal state resistivity decrease substantially. However, the break in the resistivity curve below 110 K is indicative of the emergence of Bi-2223 second phase. However, signatures of the high-$T_c$ phase in the same sample were not detected in the magnetization measurement (see Figure 5a), and by XRD scan shown in Figure 3a. We speculate that interlayer growth of extra Ca/CuO layers may occur in Bi-2212 films without the formation of the compact Bi-2223 phase. The presence of Ca/CuO inter-growth has been reported previously in LPE-grown Bi-2212 films [25] and in a series of bulk crystals showing a gradual phase transformation from the Bi-2212 to the Bi-2223 phase [30, 31].

The behavior of intrinsic Josephson junctions and Josephson plasma waves is most conveniently probed in *c*-axis transport measurements [1]. We have fabricated mesa structures (with size of 20 × 20 μm$^2$) in the three-terminal configuration on the annealed Bi-2212 film BF-3 (see the inset of Figure 6a) using photo-lithography and laser-writing techniques. The mesa height is approximately 100 nm corresponding to a stack of ~60 intrinsic Josephson junctions. The temperature dependence of the *c*-axis resistance of the mesa is shown in Figure 6a. The well-known exponential increase with decreasing temperature and a sharp transition around 82 K are clearly seen. The measured resistivity of the superconducting state does not go to zero due to contact resistance between the top gold electrode and the Bi-2212 film and/or due to deteriorated junctions at the top of the stack. Figure 6b shows current-voltage (*I-V*) characteristics at 25 K of the IJJ stack in the mesa. The *I-V* curve clearly displays the hysteretic branch structure due to resistive switching of the under-damped intrinsic Josephson junctions in the Bi-2212 film. The resistive switching of the IJJs occurs at two distinct values of the critical current ($I_c$) because part of the mesa was inadvertently patterned with half the cross-sectional area (see inset in figure 6a). The critical current densities ($J_c$) of the two parts are similar and approximately 600 A/cm$^2$. The voltage jump between the first resistive branches is approximately 14 mV and decreases gradually as an increasing number of IJJs are switched to the resistive state by the increasing bias current, due to heating effects or non-equilibrium tunneling effects [32]. The values of $T_c$, $J_c$ and the branch spacing $\Delta V_{jump}$ agree well with values of slightly underdoped Bi-2212 bulk single crystals [33].

## 4. Conclusions

We have demonstrated experimentally that high-quality and thick epitaxial Bi-2212 films can be grown with the LPE technique from a Bi-2223 melt composition at a growth temperature of ~870°C on MgO substrates. XRD and EDS analyses of our best films confirm phase-pure highly *c*-axis aligned crystalline Bi-2212 films. Transport measurements on annealed films demonstrate sharp superconducting transitions at approximately 82 K and strongly hysteretic *I-V* characteristics with relatively high $J_c$ of ~600 A/cm$^2$. The structural and superconducting properties of the films were found to be very sensitive to a large number of growth parameters in the LPE system such as growth temperature and temperature gradient, and precursor composition. These affect the mechanism of the chemical transport from the precursor composite in the KCl liquid to the substrate, and the formation of the stoichiometric crystalline Bi-2212 phase on the substrate. The observation of branched *I-V* characteristics on a mesa patterned onto an LPE-grown Bi-2212 film is an encouraging result that indicates that these thick films could potentially sustain integrated JPW devices, leading to THz-on–a-chip systems.

## Acknowledgement

We would like to thank Prof. G. Balestrino for helpful discussions on high-$T_c$ film growth by LPE. We also thank Matthew P. Smylie and Kristin Willa for their help with magneto-transport measurements. Magnetization, transport measurements were supported by the U.S. Department of Energy, Office of Science, Materials Sciences and Engineering Division, and the use of the Center for Nanoscale Materials, Office of Science user facilities, was supported by the U. S. Department of Energy, Office of Science, Office of Basic Energy Sciences.  Synthesis and related structural and compositional characterization was supported by the Laboratory Directed Research and Development (LDRD) funding from Argonne National Laboratory, provided by the Director, Office of Science, of the U.S. Department of Energy under Contract No. DE-AC02-06CH11357.

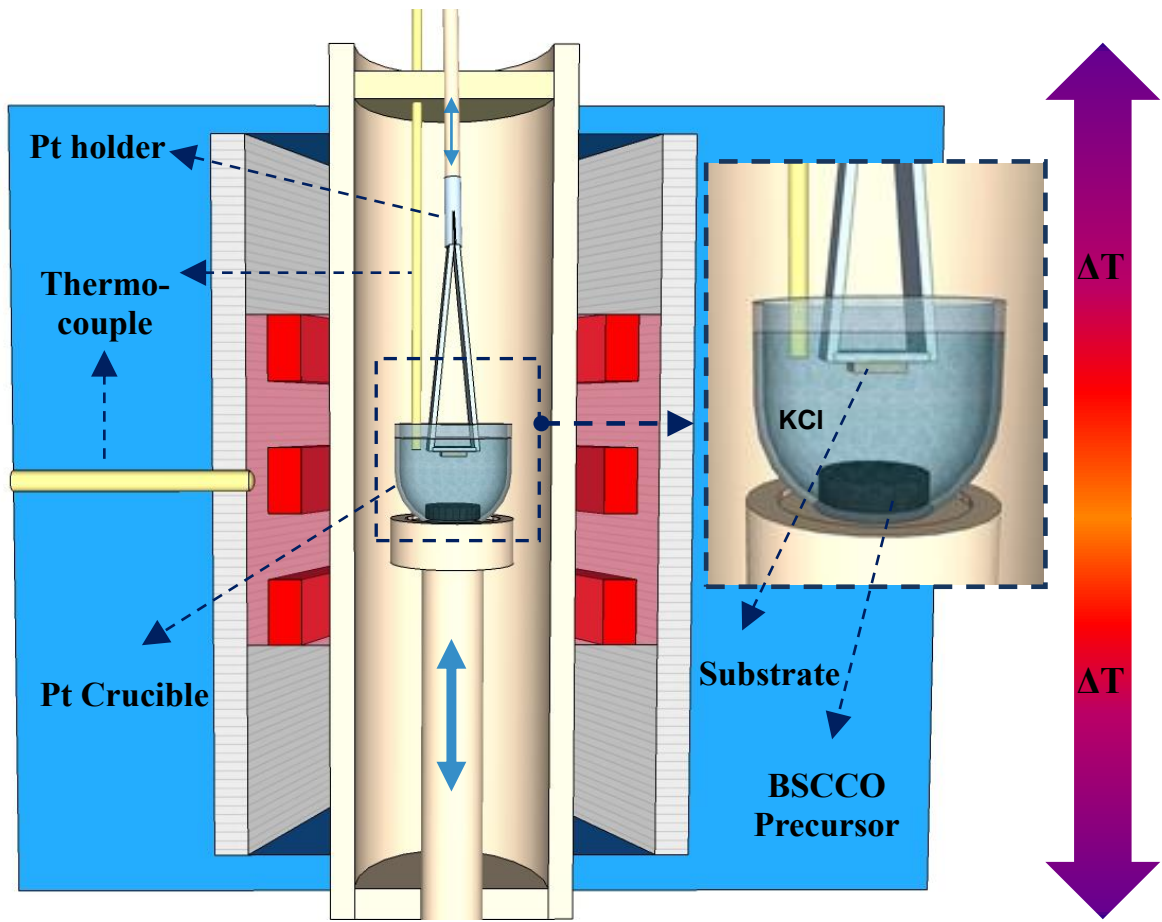

**Figure 1.** Schematic illustration of the LPE growth setup consisting of a vertical furnace, Pt crucible, alumina support, substrate holder and two thermocouples. The bottom of the crucible is aligned with the center of heat zone in the furnace. The right inset shows the position of the substrate holder and the *R*-type thermocouple inserted into the KCl liquid during the growth process.

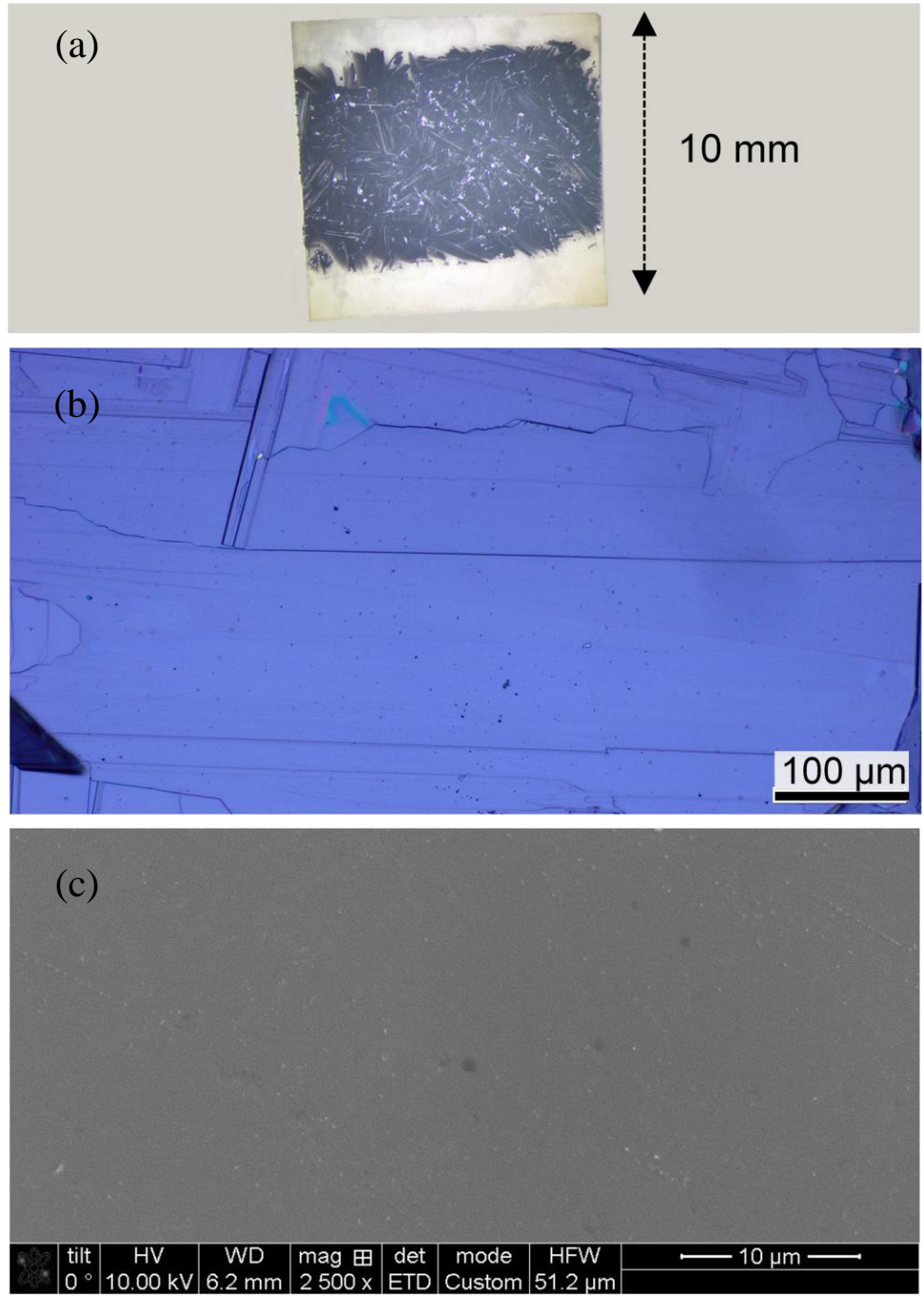

**Figure 2.** (a) Full image of an as-grown Bi-2212 film on MgO substrate with size of $1 \times 1$ cm$^2$. (b) Polarized light image showing large crystalline grains in the *a-b* plane. (c) SEM image of the film surface.

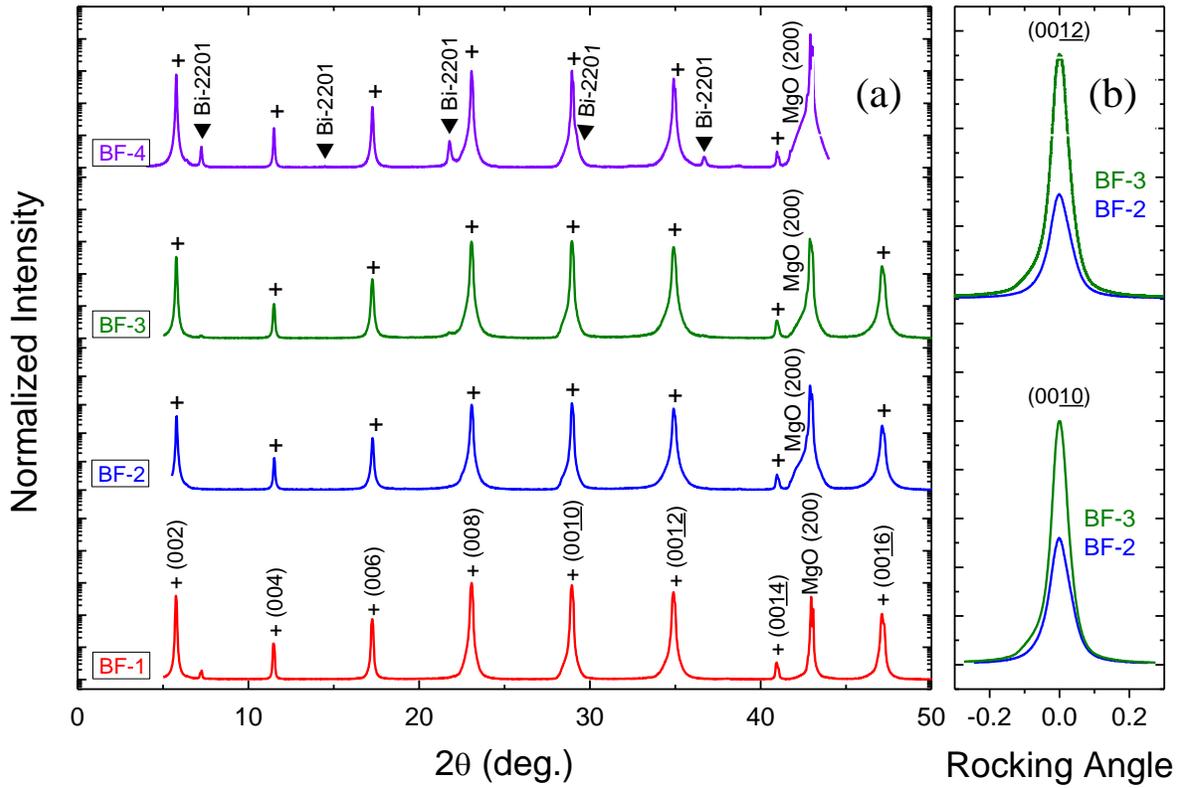

**Figure 3.** (a) XRD scans of (00 ℓ) reflections of four films sequentially grown from a Bi-2223 melt on MgO substrates showing well-orientated epitaxial grains on the substrate. Diffraction peaks from Bi-2212 and Bi-2201 phases are indicated by symbols (+ and ▼, respectively). XRD intensities of the films are plotted on a log-scale to reveal the weak diffraction peaks from crystallographic planes of the Bi-2201 phase. (b) Rocking curves of two films acquired around the (00$\underline{10}$) and (00$\underline{12}$) reflections show single sharp peaks with a full width at half maximum of less than 0.1°.

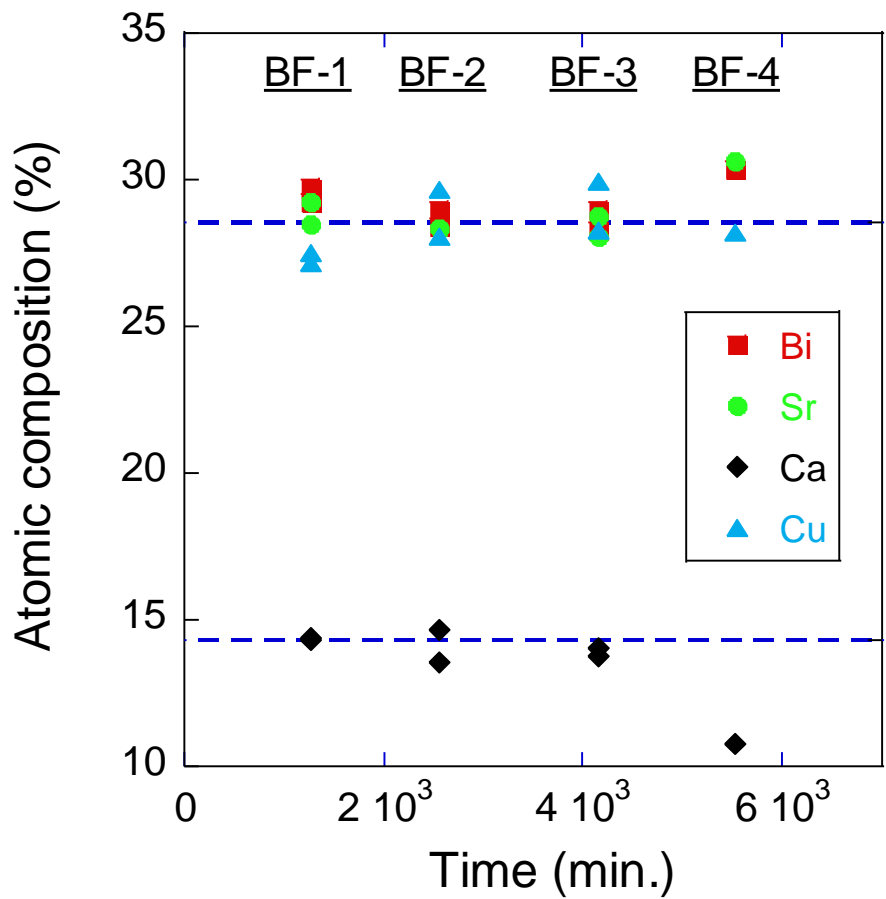

**Figure 4.** Atomic composition of the films sequentially grown from the same Bi-2223 melt derived from EDS data. Dashed lines correspond to the nominal stoichiometric composition of Bi-2212 single phase.

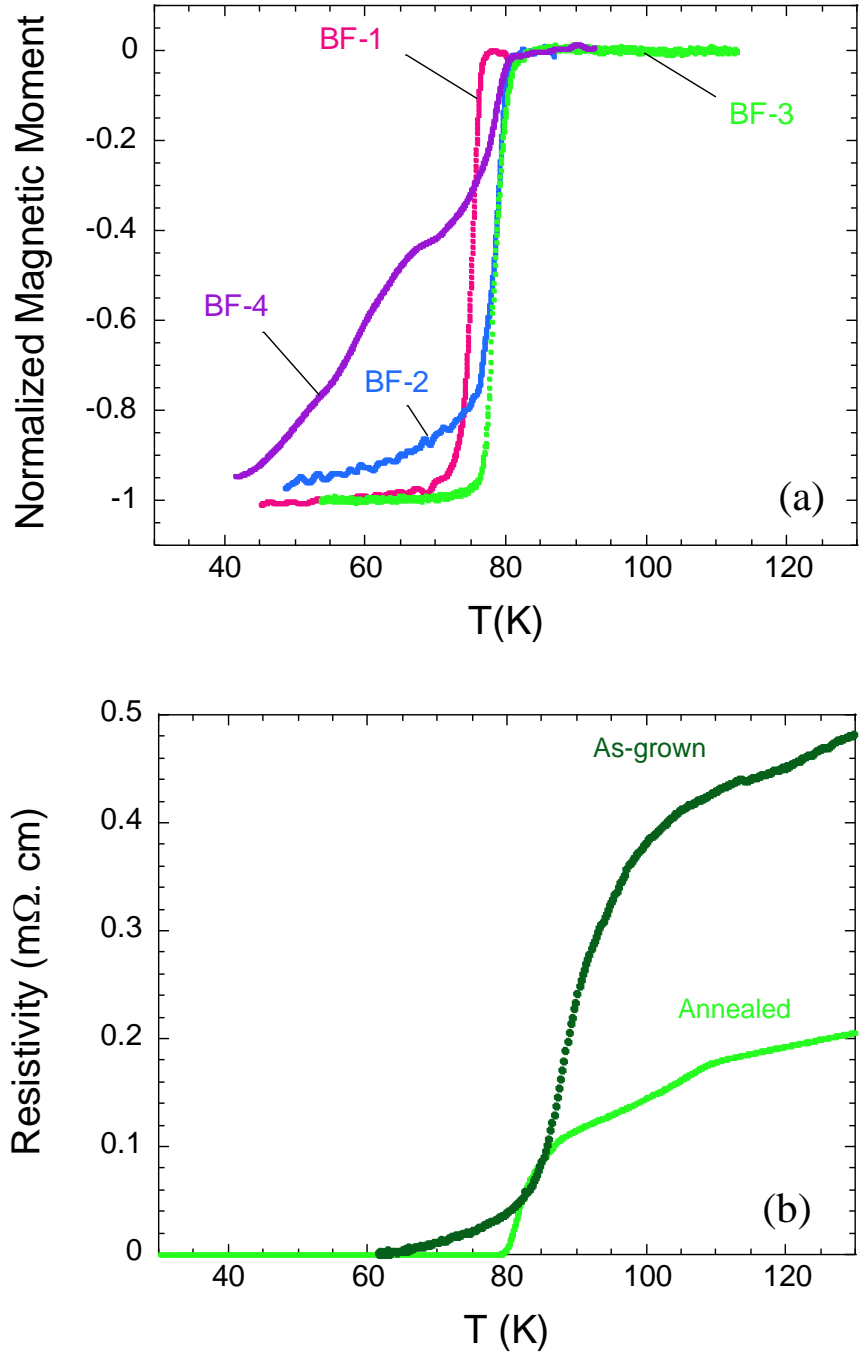

**Figure 5.** (a) Temperature dependence of the magnetic moment of the annealed Bi-2212 films. While the first three films display sharp transitions at 76 K and 80 K, the films grown at the end of the process show two wide transitions. (b) Temperature dependence of in-plane resistivity of film BF-3 before and after oxygen annealing shows two superconducting transitions clearly seen on the *R-T* curve of the annealed film.

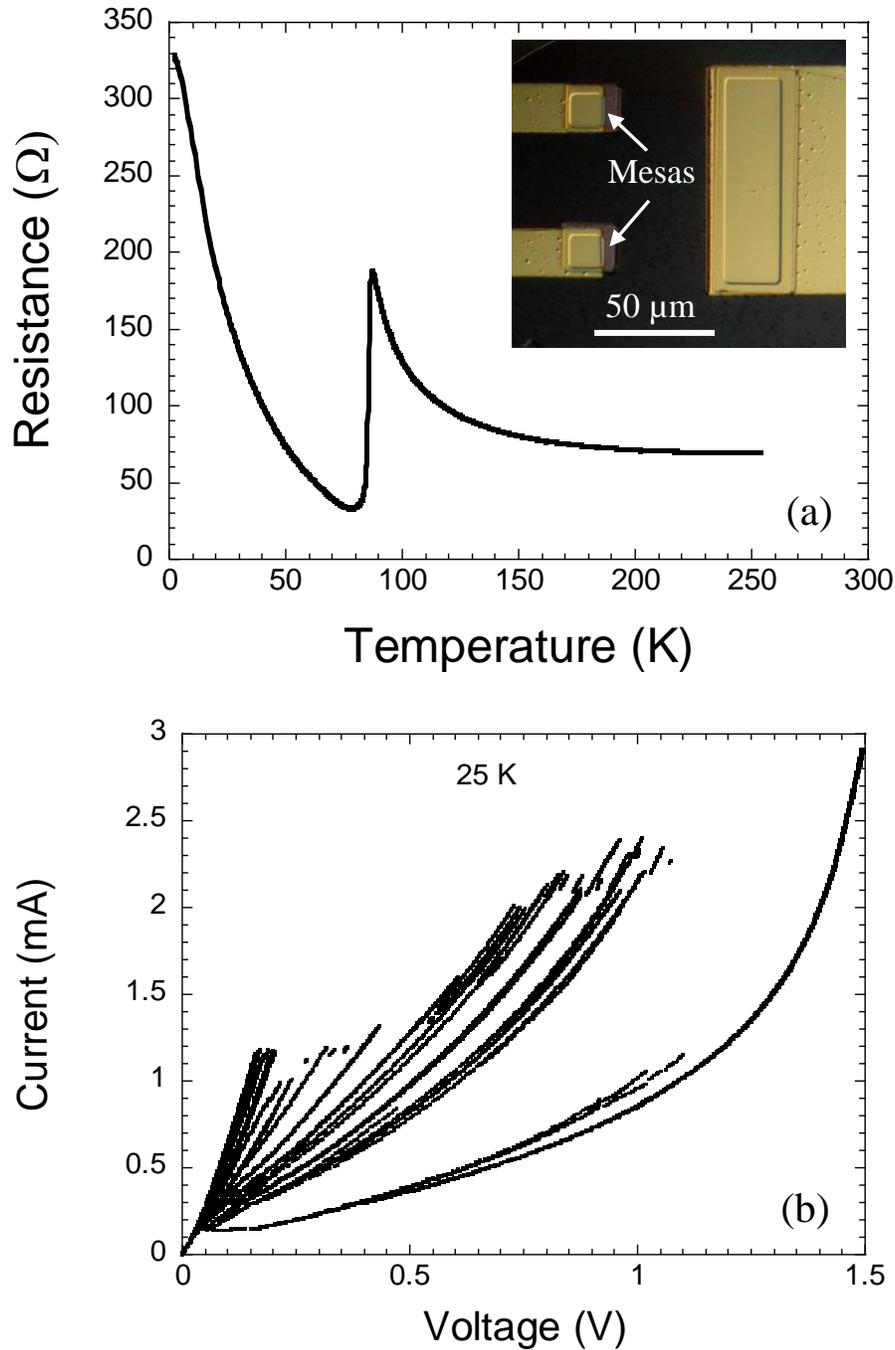

**Figure 6.** (a) Temperature dependence of resistance of a small mesa with size of 20 × 20 μm$^2$ fabricated on an annealed BSCCO film (BF-3). The inset shows an optical picture of a chip with two mesas; resistivity and *I-V* data were obtained on the top one. (b) The *c*-axis *I-V* characterization of the mesa carried out at 25 K. Both *R-T* and *I-V* measurements were performed in three-terminal configuration.